\documentclass[intlimits,twoside,a4paper]{article}

\usepackage{amsmath}
\usepackage{amssymb}
\usepackage{graphicx}
\usepackage[T2A]{fontenc}
\usepackage[cp1251]{inputenc}

\usepackage{caption}
\usepackage{subcaption}

\usepackage{color}

\usepackage{textcomp}

\usepackage{cmpj2}

\issue{2015}{18}{2}{23001}
\doinumber{10.5488/CMP.18.23001}

\title[Anomalous Brownian motion in a nematic medium]{Anomalous Brownian motion of colloidal particle in a nematic environment: effect of the director fluctuations} \label{title}%

\author[T. Turiv, A. Brodin, V.G. Nazarenko]{T. Turiv\refaddr{label1, label2}, A. Brodin\refaddr{label1, label3}, V.G. Nazarenko\refaddr{label1}} %
\addresses{
\addr{label1} Institute of Physics of the National Academy of
Sciences of Ukraine, 46 Nauky Ave., 03028 Kyiv, Ukraine
\addr{label2} Liquid Crystal Institute, Kent State University, 1425 University Esplanade, 44242 Kent, OH, USA
\addr{label3} National  Technical University of Ukraine ``Kyiv Polytechnic Institute'', 37 Peremogy Ave.,  03056 Kyiv, Ukraine
}%

\date{Received October 1, 2014, in final form December 3, 2014}

\begin{document}

\maketitle

\begin{abstract} \label{abstract}
As recently reported [Turiv~T. et al., Science, 2013, \textbf{342}, 1351], fluctuations in the orientation of the liquid crystal (LC) director can transfer momentum from the LC to a colloid, such that the diffusion of the colloid becomes anomalous on a short time scale. Using video microscopy and single particle tracking, we investigate random thermal motion of colloidal particles in a nematic  liquid crystal for the time scales shorter than the expected time of director fluctuations. At long times, compared to the characteristic time of the nematic director relaxation we observe typical anisotropic Brownian motion with the mean square displacement (MSD) linear in time $\tau $ and inversly proportional to the effective viscosity of the nematic medium. At shorter times, however, the dynamics is markedly nonlinear with MSD growing more slowly (subdiffusion) or faster (superdiffusion) than $\tau $. These results are discussed in the context of coupling of colloidal particle's dynamics to the director fluctuation dynamics.%
\pacs 05.40.Jc, 61.30.-v
\end{abstract}

\section{Introduction} \label{introduction}
Unlike the visible world around us, the atomic, molecular and small particle worlds are in a state of constant motion. This motion is widely occurring in nature and plays important role in physics, chemistry, biology and engineering \cite{Coffey96}. The physical approach to this motion (Brownian motion \cite{Brown28}), developed early in the 20th century by Einstein \cite{Einstein0506,Einstein0506a}, Smoluchowski \cite{Smoluchowski06}, and Langevin \cite{Langevin08}, still forms the basis for our understanding of these stochastic dynamics, with the main result, derived by Einstein \cite{Einstein0506,Einstein0506a}, being that the mean squared displacement (MSD) $\left\langle \Delta \mathbf{r}^{2} \right\rangle $ of a particle undergoing Brownian motion in a Newtonian fluid increases linearly with time, $\left\langle \Delta \mathbf{r}^{2} (\tau )\right\rangle =6D\tau $, where $D$ is the diffusion constant. For a spherical particle with hydrodynamic radius $R$ in a fluid with viscosity $\eta $, the diffusion coefficient is given by the Stokes-Einstein relation $D=k_\textrm{B} T/\zeta $, where $k_\textrm{B} $ is the Boltzmann constant, $T$ is the temperature and $\zeta $ is the viscous friction coefficient, which under no-slip conditions is given by $\zeta =6\pi R\eta $. The normal, linear diffusion regime for a particle of mass $m$ is established on time scales long compared to the microscopic time $m/\zeta $, which is typically in the nanosecond range. These results are valid for Brownian motion under the influence of two forces, the viscous frictional force linear in particle velocity, $\mathbf{F}=-\zeta \mathbf{v}$, and a random force with white-noise-spectrum due to random collisions with surrounding particles.%

Brownian particles in complex systems may, however, exhibit quite different dynamics, reflecting the properties of particles themselves (e.g., the effects of a crossover from short-time anisotropic to long-time isotropic diffusion for ellipsoidal particles in water \cite{Han06}) as well as the local properties of the host medium that may be inhomogeneous, exhibit nonlinear friction or elastic properties, etc., with $\langle \Delta \mathbf{r}^{2} (\tau )\rangle \propto \tau ^{\alpha } $, where the exponent $\alpha $ may be different from 1. For example, colloidal particles in polymer networks \cite{Sprakel07,Sprakel08}, in F-actin networks \cite{Wong04}, in surfactant formed lyotropic liquid crystal \cite{Alam11} may exhibit a subdiffusive behavior with $0<\alpha <1$; superdiffusion, $\alpha >1$, was observed in concentrated suspensions of swimming bacteria \cite{Wu01,Wu01a} and in ``living polymers'' \cite{Ott90}.%

The behaviour of colloidal particles in a nematic liquid crystal (NLC) is in many respects more complicated as compared to the isotropic fluid host. First of all, the particle sets a certain director distortion around itself, due to the anisotropic nature of surface interactions (surface anchoring) \cite{Poulin97}. These director distortions lead to long-range elastic interactions of the particle with the bounding walls. Second, the orientational nematic order leads to an anisotropy of the Stokes drag \cite{Ruhnwald96,Abras12,Skarabot10,Moreno-Razo11,Mondiot12,Stark01,Loudet04,Koenig09,Stark02,Stark03}. As a result, Brownian motion in a nematic host becomes anisotropic with two different diffusion constants $D_{\parallel} $ and $D_{\bot}$, corresponding to the directions parallel and perpendicular to the nematic director $\mathbf{n}$. Third, although the LC medium is homogeneous, the average axes of orientation fluctuate in time and space and thus influence the diffusive regimes \cite{Turiv13,Park14}. In addition to normal diffusion, the particle experiences two anomalous regimes, with MSD growing slower (subdiffusion) and faster (superdiffusion) than $\tau $. The anomalous diffusion occurs at time scales that correspond to the relaxation times of director fluctuations \cite{Turiv13}. All three regimes of diffusion are anisotropic, with the MSD being larger for the motion along the director. Once the nematic is melted, the diffusion becomes normal and isotropic.%

In this paper, we elaborate in detail the Brownian dynamics of colloidal particles in a nematic host on time scales shorter than the expected time of director fluctuations $\tau _\textrm{relax} $. Following the experiments described in \cite{Turiv13}, we first give a detailed description of the particle tracking experiments with an extensive analysis of the possible errors in the observed trajectories of the colloids. This is followed by an analysis of the diffusion of colloids, which turns out to be nonlinear with time for a certain time lag, as it is described in \cite{Turiv13}. In the theoretical section, we present a model consideration of the diffusion of the colloidal particle in LC, where we explain the anomalous regime by coupling of the particle motion and director fluctuations.%

\section{Experiment} \label{experiment}

\subsection{Material} \label{material}

The refractive indices for ordinary ($n_{\mathrm{o}} $) and extraordinary ($n_{\mathrm{e}} $) modes of light propagation in a nematic are different from each other. The director $\mathbf{n} $ is also the local optic axis. Whenever the director fluctuates, birefringence $\Delta n=n_{\mathrm{e}} -n_{\mathrm{o}} $ and the associated ``lens'' effect of the distorted optic axis around the particle translate these fluctuations into phantom drifts of the image. %
\begin{figure}[h!]
\centering
\includegraphics[scale=0.3, trim=0 10 0 60, clip]{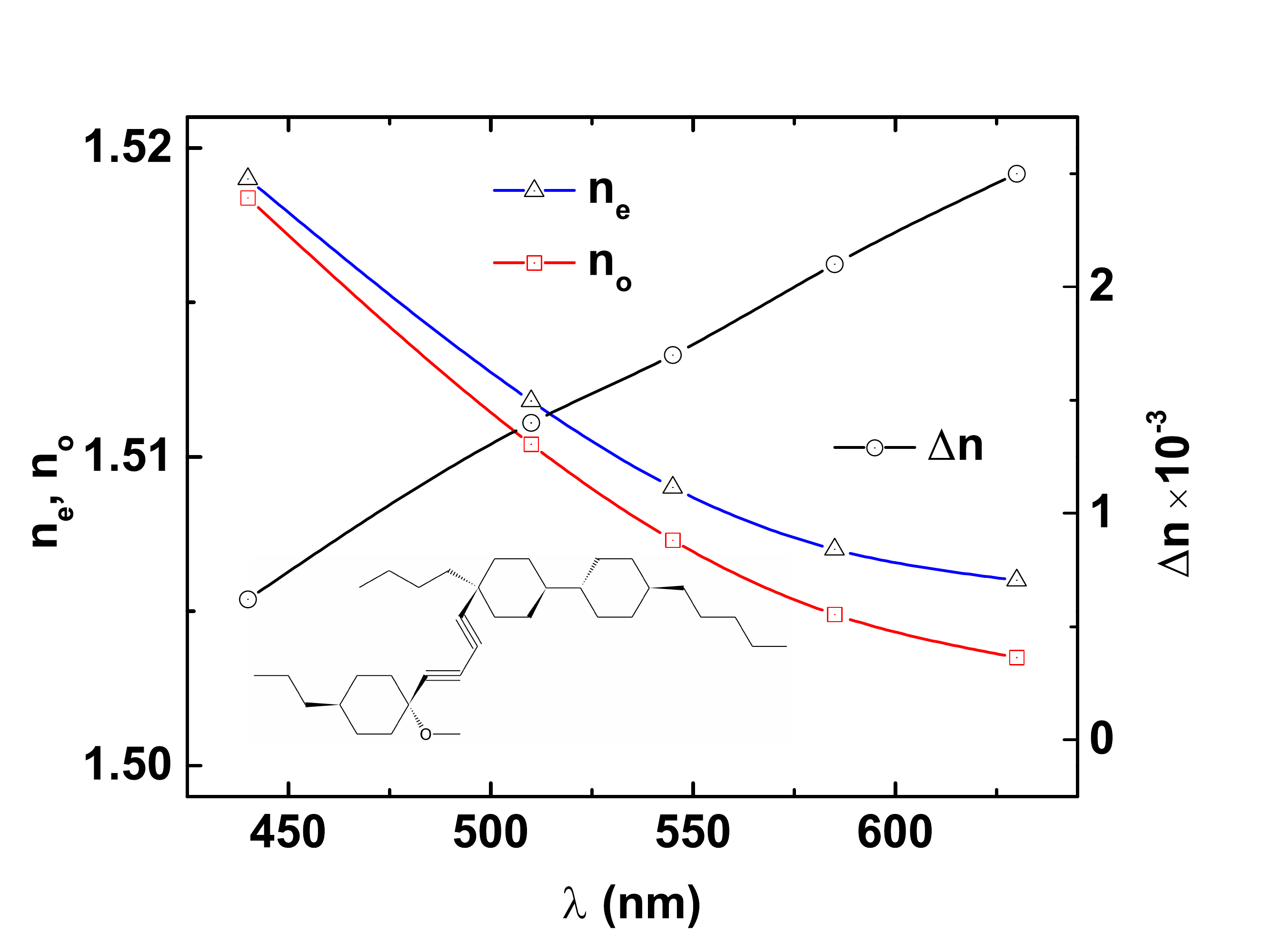}
\caption{(Color online) Wavelength dependence of the refractive indices and optical anisotropy of the NLC IS--8200 which chemical structure formula is placed in the plot.}
\label{fig1}
\end{figure}%
To overcome the problem, we use the nematic IS--8200 synthesized at Merck (Germany) with ultra-low birefringence $\Delta n=0.0015$ (at light wavelength $\lambda =520$~{nm}) \cite{Reiffenrath01}. IS--8200 is a thermotropic (solvent free, single-component) material with the nematic state in the temperature range 47--53\textcelsius, above which it melts into the isotropic fluid, figure~\ref{fig1}. Small $\Delta n$ suppresses the image shifts caused by fluctuations.%

\subsection{Director alignment} \label{director_alignment}

\begin{figure}[!h]
\hspace{1.5cm}
\includegraphics[width=0.3\textwidth]{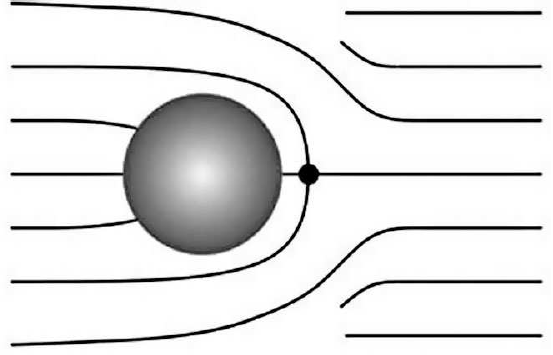}%
\hfill%
\includegraphics[width=0.3\textwidth]{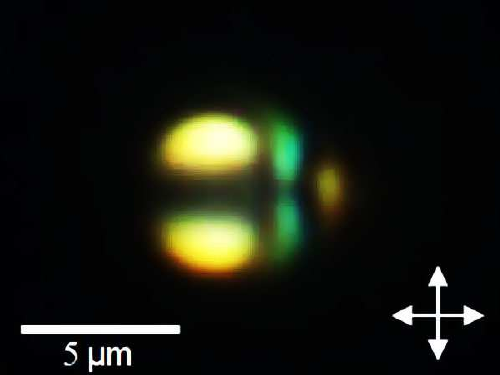}%
\hspace{1.5cm}
\\%
\parbox[t]{0.4\textwidth}{%
\centerline{\hspace{1.5cm}(a)}%
}%
\parbox[t]{0.4\textwidth}{%
\centerline{\hspace{4.5cm}(b)}%
}
\caption{(Color online) (a) Schematic director field around a sphere with normal surface anchoring  and (b) optical image of 5~\textmu{}m silica sphere of the dipole structure immersed in IS--8200 placed between crossed polarizers.}
	\label{fig2}
\end{figure}

Alignment of the director at both the particle's surface and the bounding plates influences the diffusion and thus needs to be controlled. We explored the alignment at the surface of spheres to be perpendicular to the surface. In his case, the particles were functionalized with dimethyl-octadecyl- [3(trimethoxysilyl)propyl] ammonium chloride (DMOAP) \cite{Skarabot05}. The overall uniform orientation of the nematic was set by two glass plates covered with rubbed polyimide PI--2555 (Nissan Chemicals) alignment layers that produce a uniaxial planar alignment $\textbf{n} _{0} =(1,0,0)=\textrm{const} $ in the cell. The locally distorted director around the spheres should smoothly transform into $\textbf{n} _{0} $ \cite{Loudet04}. The resulting equilibrium director configuration is of a dipolar type (with a point defect~--- hyperbolic hedgehog accompanying the sphere, figure~\ref{fig2}) for the normal anchoring. The director distortions around the particle \cite{Lubensky98} lead to a repulsion from the bounding substrates \cite{Pishnyak07}. The particles levitate \cite{deGennes93} in the bulk at some height determined by the balance of gravity and elastic forces.%
\begin{figure}[!h]
\includegraphics[width=0.48\textwidth]{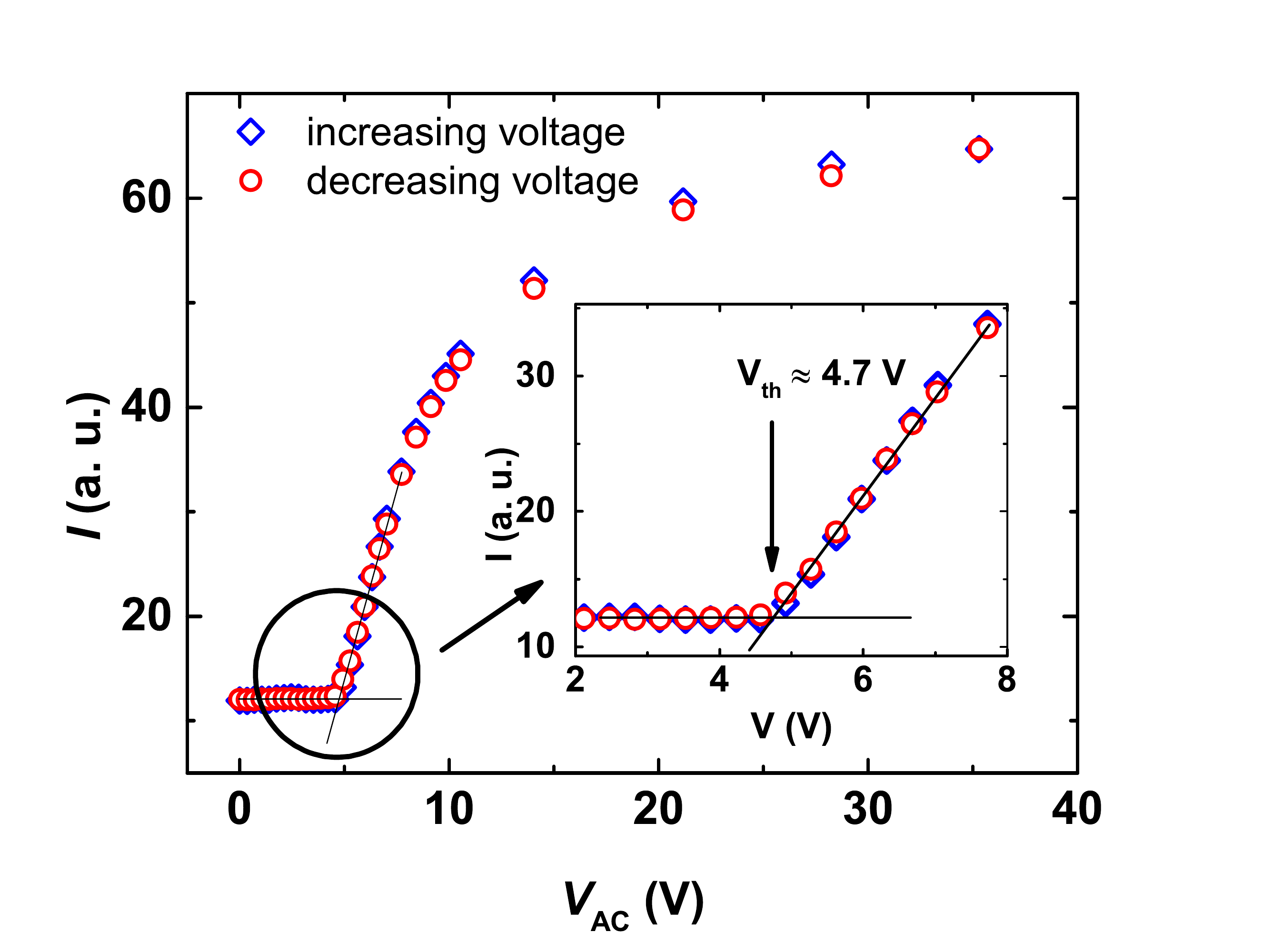}%
\hfill%
\includegraphics[width=0.48\textwidth]{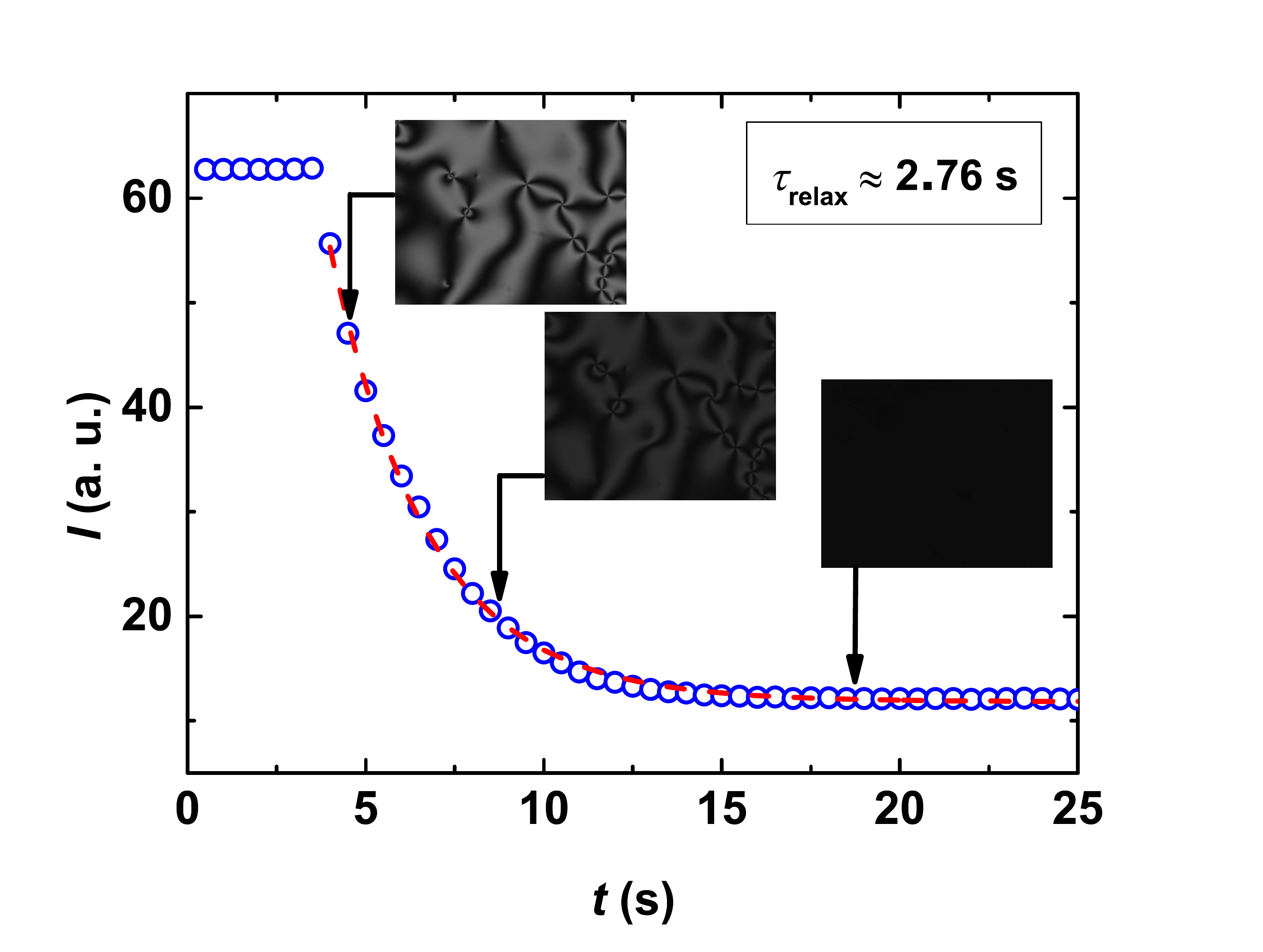}%
\\%
\parbox[t]{0.4\textwidth}{%
\centerline{(a)}%
}%
\hfill
\parbox[t]{0.4\textwidth}{%
\centerline{(b)}%
}
\caption{(Color online) (a) Intensity of the transmitted light as a function of the applied voltage. (b) Intensity of the transmitted light as a function of time after voltage is switched off. IS--8200 liquid crystal, homeotropic alignment, crossed polarizers. Cell thickness 20~\textmu{}m.}
	\label{fig3}
\end{figure}

\subsection{Time scale} \label{time_scale}

The director field $\mathbf{n}\left(\mathbf{r},t\right)$ fluctuates in time and space \cite{deGennes93}. The relaxation time of fluctuations is much slower than the relaxation time of vorticity  $\sim \rho l^{2} /\eta \sim 1$~\textmu{}s, which is the time needed by the perturbed fluid of density $\rho $ and viscosity $\eta \sim 0.1~\textrm{Pa}\cdot \textrm{s}$ to flow over the distance $l$. For a director perturbation with a wavevector $\mathbf{q}$, this time is $\tau _\textrm{relax} \sim \eta _\textrm{eff} /(q^{2} K)$, where $\eta _\textrm{eff} $ and $K$ are the effective viscosity and elastic constant. A uniaxial nematic features five independent viscosities and at least four Frank elastic constants, thus $\eta _\textrm{eff} $ and $K$ are complex combinations of these that depend on the director field \cite{Kleman03}. In a flat cell of a finite thickness $h$, the fluctuation spectrum is restricted; in case of strong director anchoring at the boundaries, the minimum wavevector component in the $z$-direction perpendicular to the cell boundaries is $q_{\min }^{z} =\pi /h$ \cite{Vilfam02}. We are interested in wavevectors not much larger than $q=\pi /d$, where $d$ is the sphere's diameter, since perturbations with wavelength much shorter than $d$ will produce on average a negligible effect. The corresponding range of viscous relaxation times is thus $\tau _{d} <\tau _\textrm{relax} <\tau _{h} $, where $\tau_{h} =\eta _{\textrm{eff}} h^{2} /\pi ^{2} K$ and $\tau_{d} =\eta _{\textrm{eff}} d^{2} /\pi ^{2} K$. To estimate these,  we experimentally determined the quantity $\eta _{\textrm{eff}} /K$ in the splay Frederiks transition of IS--8200 and found $\eta _{\textrm{eff}} /K\approx 10^{11}$~{s/m}$^{2}$, figure~\ref{fig3}.%

\subsection{Optical particle tracking: experimental and data analysis procedures} \label{particlke_tracking}

Experimental setup consists of a CMOS high-speed video camera MotionBlitz EoSens mini1 (Microtron GmbH) mounted on an inverted microscope Nikon TE2000-U with a $100\times 1.3~N.A.$ immersion objective. The camera is capable of a 5000~fps maximum frame rate (time resolution 0.2~ms); the images are gray-scale with 256 gradations (bit depth 8). The pixel size is $14\times 14$ microns, so that the magnification is 140~nm/pixel, with the Airy disk covering $\sim 100$ pixels. For temperature control, we used a Linkam LTS120 heating stage (accuracy 0.3\textcelsius).%

For a point light source, its image is diffraction-blurred into an Airy disk, whose intensity profile in the image plane has a root-mean-square (r.m.s.) size of $s=0.44\lambda M/2~N.A.$, where $\lambda $ is the wavelength, $N.A.$ the numerical aperture, and $M$ the transverse magnification of the optical system. $N.A.$ should be maximized in order to decrease the diffraction blur. Meanwhile, since the image is pixelised as it is formed on a detector matrix, larger magnification $M$ implies more pixels in the image, and thus more information and better ultimate position accuracy. In practice, $N.A.$ and $M$ are determined by the objective lens, typically with $N.A.\sim 1$ and $M\sim 100$, so that ${s}\sim 10$~\textmu{}m.%

Particle trajectories are almost invariably analyzed in terms of mean square displacement \cite{Crocker96}, MSD, i.e., $\left\langle \Delta x^{2} (\tau )\right\rangle $ and $\left\langle \Delta y^{2} (\tau )\right\rangle $, where $\tau $ is the time lag and angular brackets stand for ensemble average. In an isotropic medium $\left\langle \Delta x^{2} (\tau )\right\rangle $ and $\left\langle \Delta y^{2} (\tau )\right\rangle $ are equal, whereas in an anisotropic medium, such as NLC, they are different. Ensemble average cannot be obtained from a single trace. Instead, one can do a time average which is believed to be the same as the ensemble average in the limit of infinite averaging time. Specifically, one computes
\begin{equation} \label{eq1}
	\left\langle \Delta x^{2} (\tau )\right\rangle =\left\langle [x(t+\tau )-x(t)]^{2} \right\rangle ,\qquad \left\langle \Delta y^{2} (\tau )\right\rangle =\left\langle [y(t+\tau )-y(t)]^{2} \right\rangle,
\end{equation}
where angular brackets now stand for time average. Even though time averages equal the ensemble averages in an ergodic system, time averages over a trace of finite length and time step possess specific statistical errors \cite{Lee84}.%

The measurement errors ($\delta x_{i}$, $\delta y_{i}$) (assumed to have zero mean) in particle coordinates ($x_{i},y_{i}$) yield a positive additive contribution to MSD computed through equation \eqref{eq1}, as it is a quadratic form. Assuming $(x_{i},y_{i}) = (\tilde{x}_{i} +\delta x_{i} ,\tilde{y}_{i} +\delta y_{i})$, where $(\tilde{x}_{i} ,\tilde{y}_{i} )$ are the true coordinates, and denoting $x_{i} =x_{t} ,\, \, \delta x_{i} =\delta x_{t} $, where $t=i\Delta t$, one gets for $\left\langle \Delta x^{2} (\tau )\right\rangle $ (and similarly for $\left\langle \Delta y^{2} (\tau )\right\rangle $)
\begin{eqnarray} \label{eq2}
	\left\langle \Delta x^{2} (\tau )\right\rangle &=&\left\langle [(\tilde{x}_{t+\tau } +\delta x_{t+\tau } )-(\tilde{x}_{t} +\delta x_{t} )]^{2} \right\rangle  \nonumber \\
&=&\left\langle (\tilde{x}_{t+\tau } -\tilde{x}_{t} )^{2} \right\rangle +\left\langle (\tilde{x}_{t+\tau } -\tilde{x}_{t} )(\delta x_{t+\tau } -\delta x_{t} )\right\rangle +\left\langle (\delta x_{t+\tau } -\delta x_{t} )^{2} \right\rangle ,
\end{eqnarray}
where the first term is the ``true'' MSD. Assuming that the errors are uncorrelated with the coordinates, $\left\langle \tilde{x}\delta x\right\rangle =0$, the second term in equation \eqref{eq2} averages to zero. Assuming further that the errors of different trace points are uncorrelated, $\left\langle \delta x_{t+\tau } \delta x_{t} \right\rangle =0$, and have the same variance $\left\langle \delta x^{2} \right\rangle =\Delta _{0}^{2} $, the last term in equation \eqref{eq2} yields $2\Delta _{0}^{2} $. Thus,
\begin{equation} \label{eq3}
	\left\langle \Delta x^{2} (\tau )\right\rangle =\left\langle \Delta \tilde{x}^{2} (\tau )\right\rangle +2\Delta _{0}^{2} \,,
\end{equation}
so that measurement errors result in a constant additive background in MSD. The assumption of uncorrelated errors almost certainly holds true for static errors, but it may not be the case for dynamic errors. For instance, the errors due to birefringence fluctuations are likely correlated to some extent over the time interval corresponding to the time scales of the fluctuations. This will add a time lag dependence to the background so that, in general, it is a function of $\tau $, i.e., $\Delta _{\tau }^{2} $. The presence of the background should be taken into account in data analyses. We will experimentally estimate these errors and their time dependence by determining the apparent MSD of immobilized particles.%
\begin{figure}
\hspace{1cm}
\includegraphics[width=0.4\textwidth]{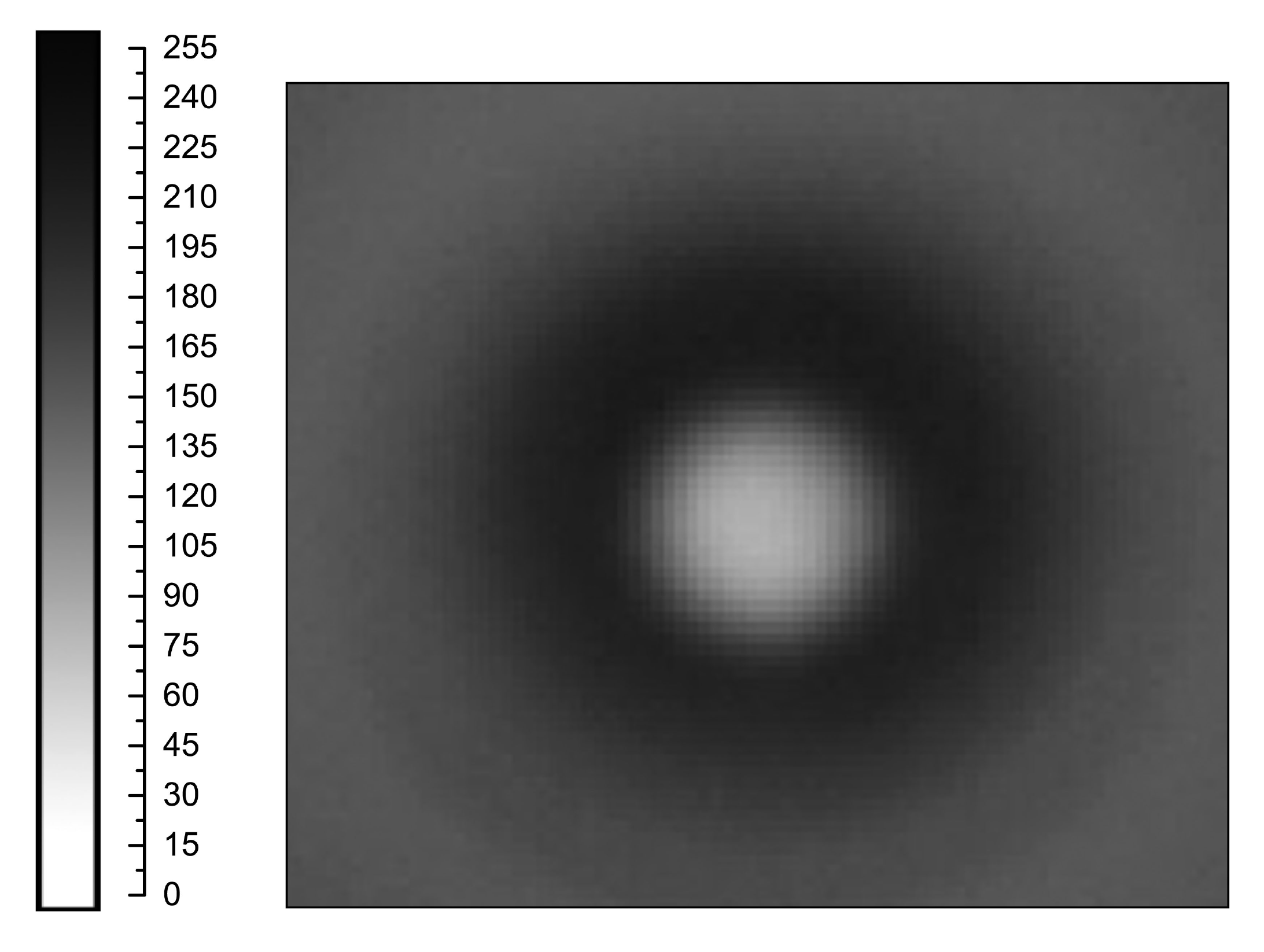}%
\hfill%
\includegraphics[width=0.37\textwidth]{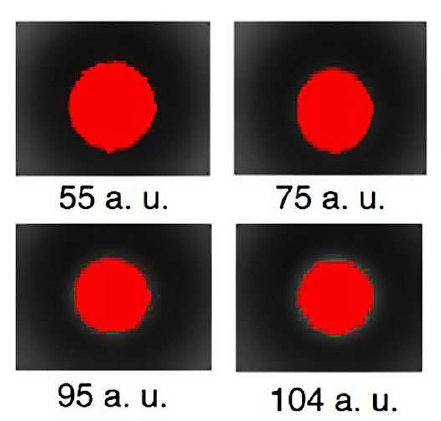}%
\hspace{1cm}
\\%
\parbox[t]{0.4\textwidth}{%
\centerline{(a)}%
}%
\hfill
\parbox[t]{0.4\textwidth}{%
\centerline{(b)}%
}
\includegraphics[width=0.48\textwidth]{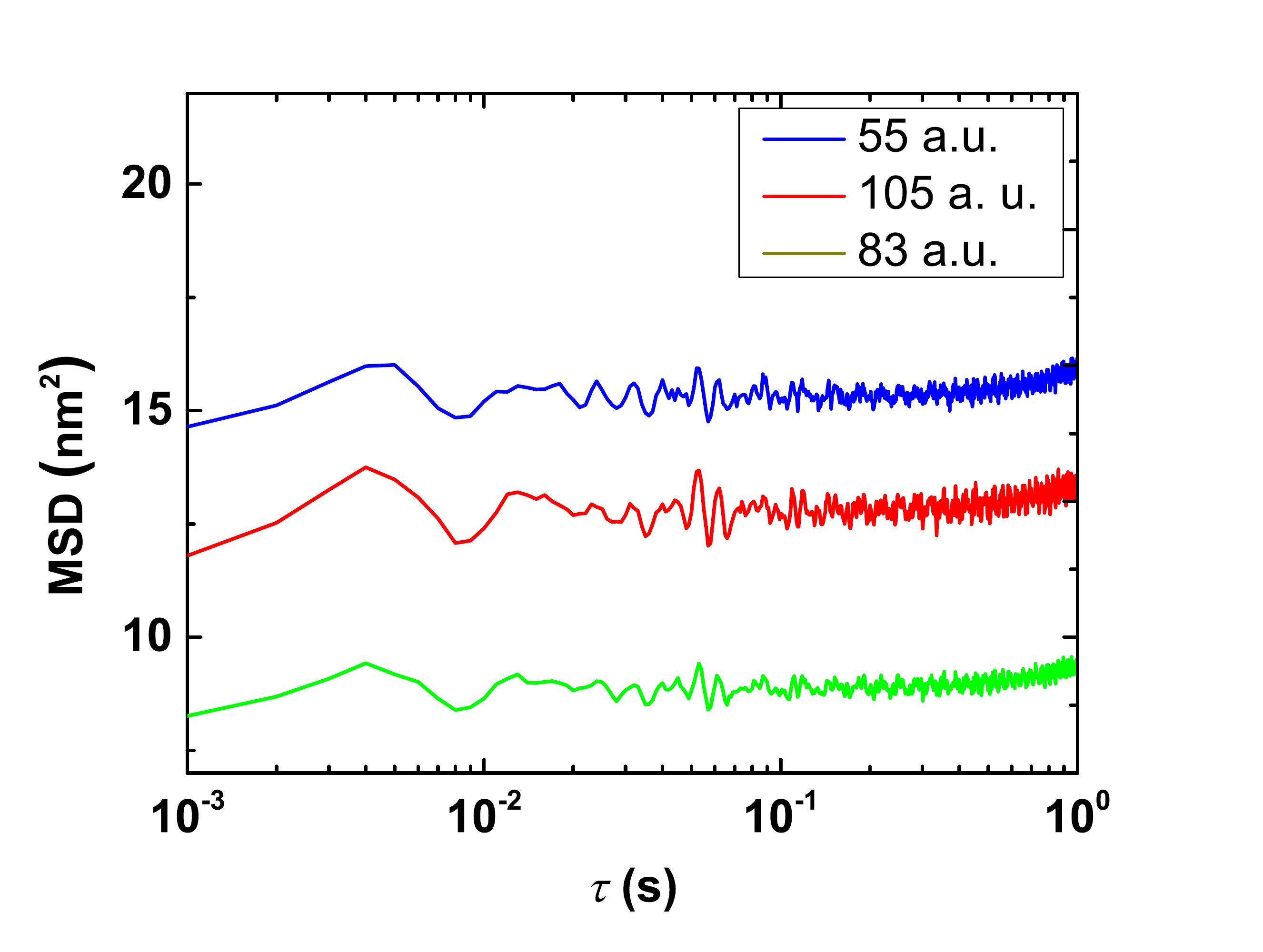}%
\hfill%
\includegraphics[width=0.48\textwidth]{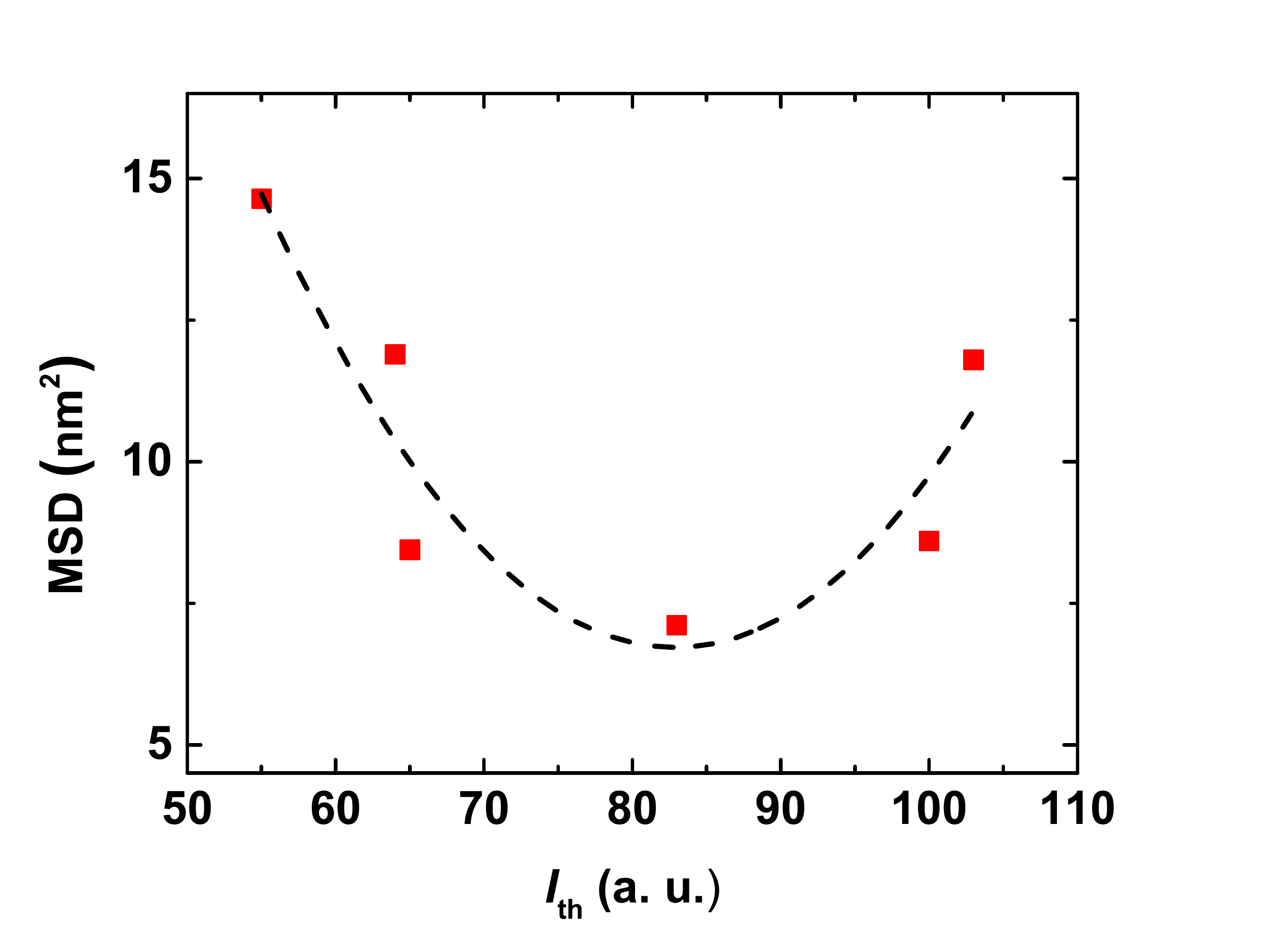}%
\\%
\parbox[t]{0.4\textwidth}{%
\centerline{(c)}%
}%
\hfill
\parbox[t]{0.4\textwidth}{%
\centerline{(d)}%
}
\caption{(Color online) (a) Grayscale image of 5~\textmu{}m particle glued to the bottom substrate of the cell. (b) Images of the particle recorded at different intensity threshold levels. Red area consists of pixels that are used to calculate the particle's center coordinates. (c) Apparent MSD of a glued particle at different levels of threshold intensity. (d) The optimum value of intensity threshold $I_{\textrm{th}}$ is determined from the minimum of MSD for the glued particle.}
	\label{fig4}
\end{figure}
Digital images of colloidal particles, captured at a maximum frame rate of 2400~fps (time resolution is 0.4~ms), were analyzed to find the coordinates ($x$, $y$) of the particle's center using the intensity-weighted algorithms. Specifically, we computed
\begin{equation} \label{eq4}
	x=\frac{\sum _{i}I_{i} x_{i}  }{\sum _{i}I_{i}  } \,,
\qquad
y=\frac{\sum _{i}I_{i} y_{i}  }{\sum _{i}I_{i}  } \,,
\end{equation}
where $x_{i} $ and $y_{i} $ are coordinates of $i$-th pixel, $I_{i} $ is its intensity. On each frame with 8-bit gray scale (0--255 arbitrary units of intensity) we take into account only the pixels that have an intensity higher than certain threshold intensity, $I_{i,\textrm{th}} $. The different threshold intensity results in different apparent MSD for an immobilized particle. Images of the particle at different threshold levels are demonstrated in figure~\ref{fig4}. Red area consists of pixels that are taken for the calculation of the particle's center coordinates. The optimal level of the threshold which gives minimum value of the apparent MSD of glued particle was extracted from the dependence of MSD vs intensity threshold, figure~\ref{fig4}.%

\begin{figure}[!h]
\centering
\includegraphics[scale=0.3, trim=0 30 0 40, clip]{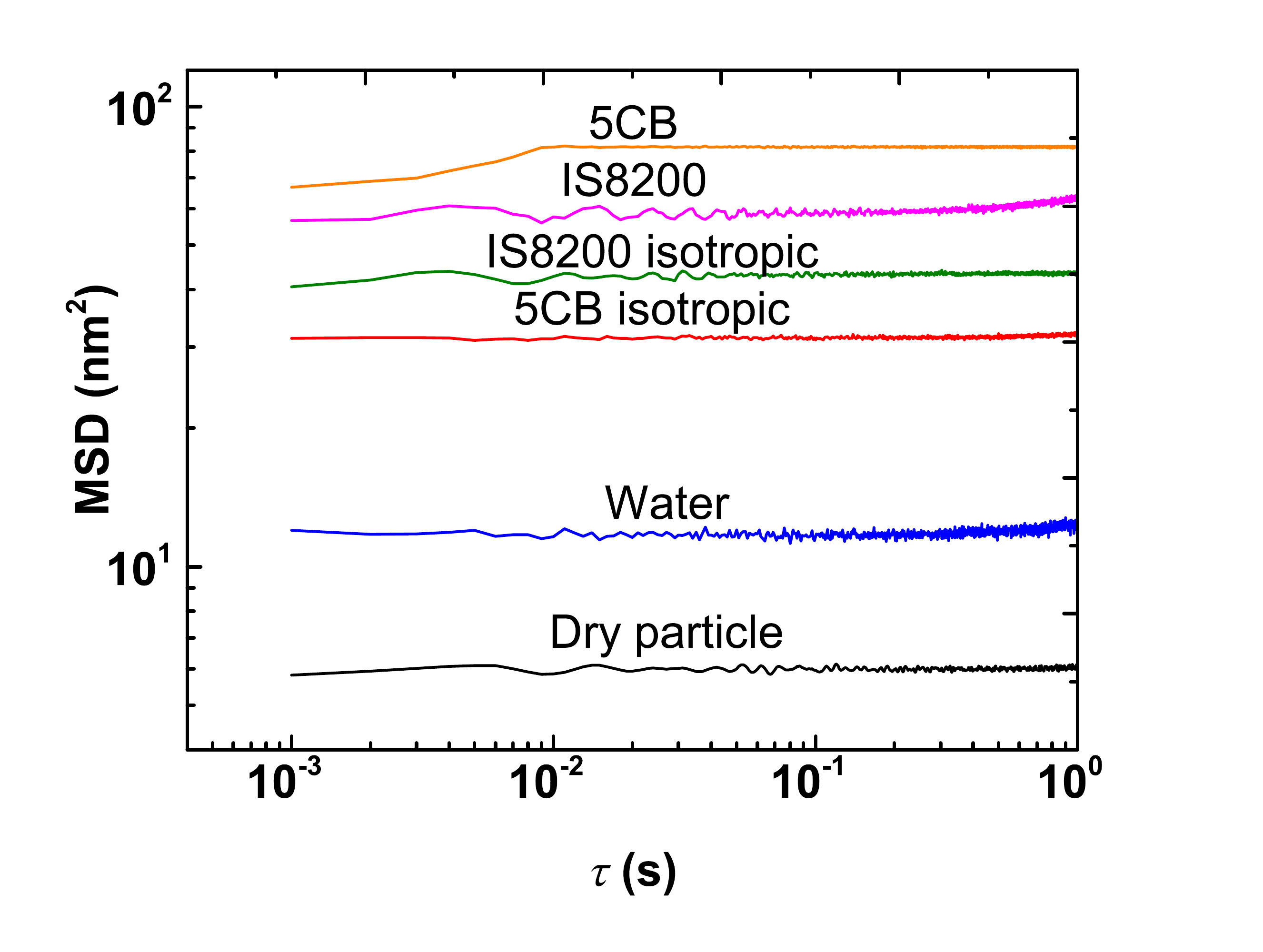}
\caption{(Color online) MSD versus time lag for immobilized silica spheres of diameter 5~\textmu{}m in four different cells (all of thickness 50~\textmu{}m) filled with two types of a NLCs and water; the data for an empty cell are labeled as ``dry particle''. Normal surface anchoring. Higher birefringence results in a higher apparent displacement of the particles.}
\label{fig5}
\end{figure}

To establish the limit of accuracy in the measurements of particle's positions that depends on birefringence, we used particles immobilized (by a Norland adhesive) at the bottom plate of the cell filled with three different fluids: two types of a nematic and water as isotropic fluid. The probing light beam traveled twice through the entire thickness of the cell. The apparent mean square displacement of the immobilized particles vs time lag is shown in figure~\ref{fig5}. The apparent displacements represent a cumulative effect of errors in measuring the particle's position caused by the optical system of the microscope, vibrations and birefringence. It grows with birefringence of the material, being the largest for the nematic pentylcyanobiphenyl (5CB) with the highest birefringence ($\sim 0.2$). In all cases, the apparent MSD in the time range of interest was about $10^{-16}$~{m}$^{2}$ or less; these values are about 100 times smaller than the MSD of free spheres experiencing on these timescales an anomalous diffusion described in the main text.%

\section{Results} \label{results}%

\begin{figure}
\includegraphics[width=0.48\textwidth]{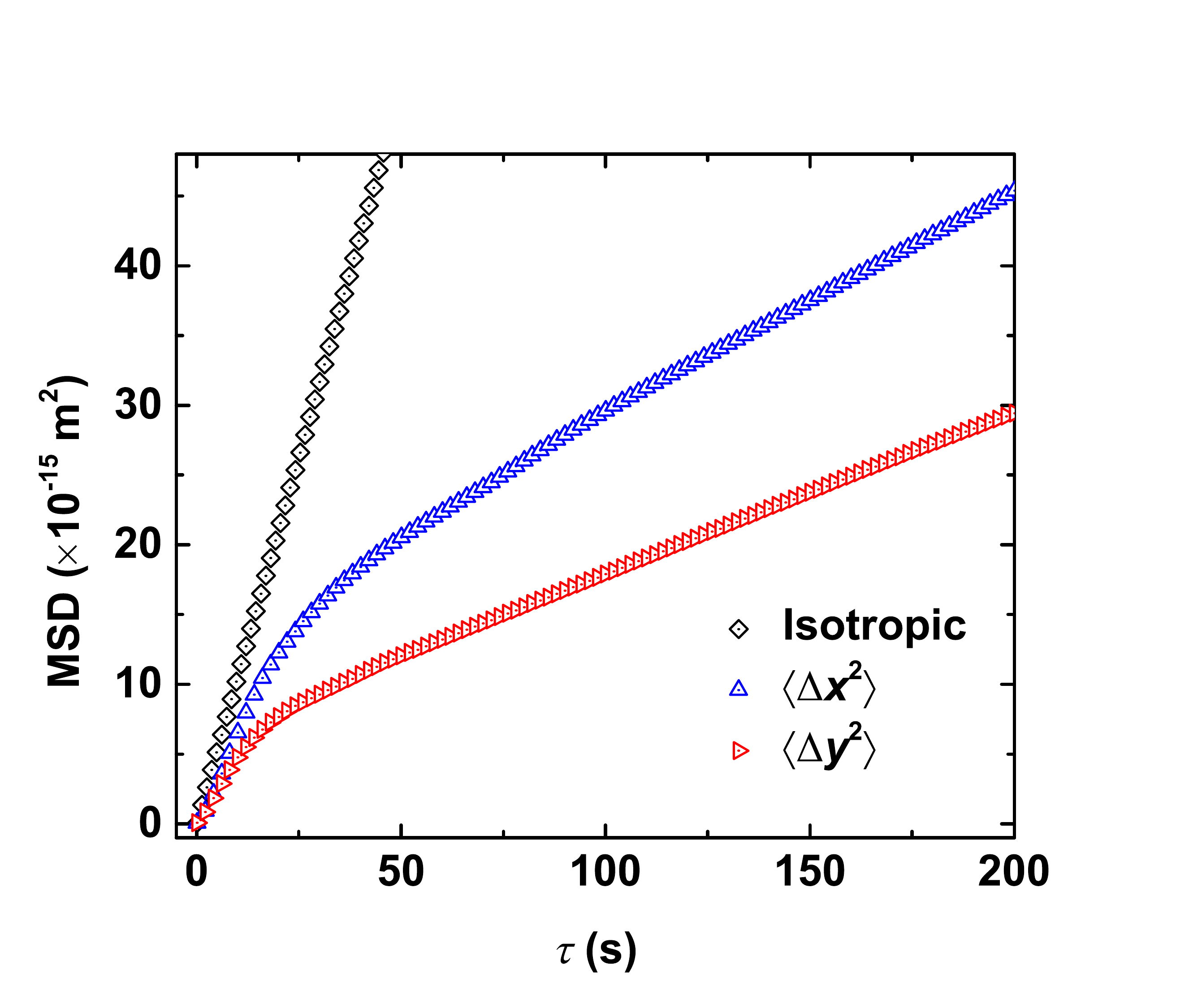}%
\hfill%
\includegraphics[width=0.48\textwidth]{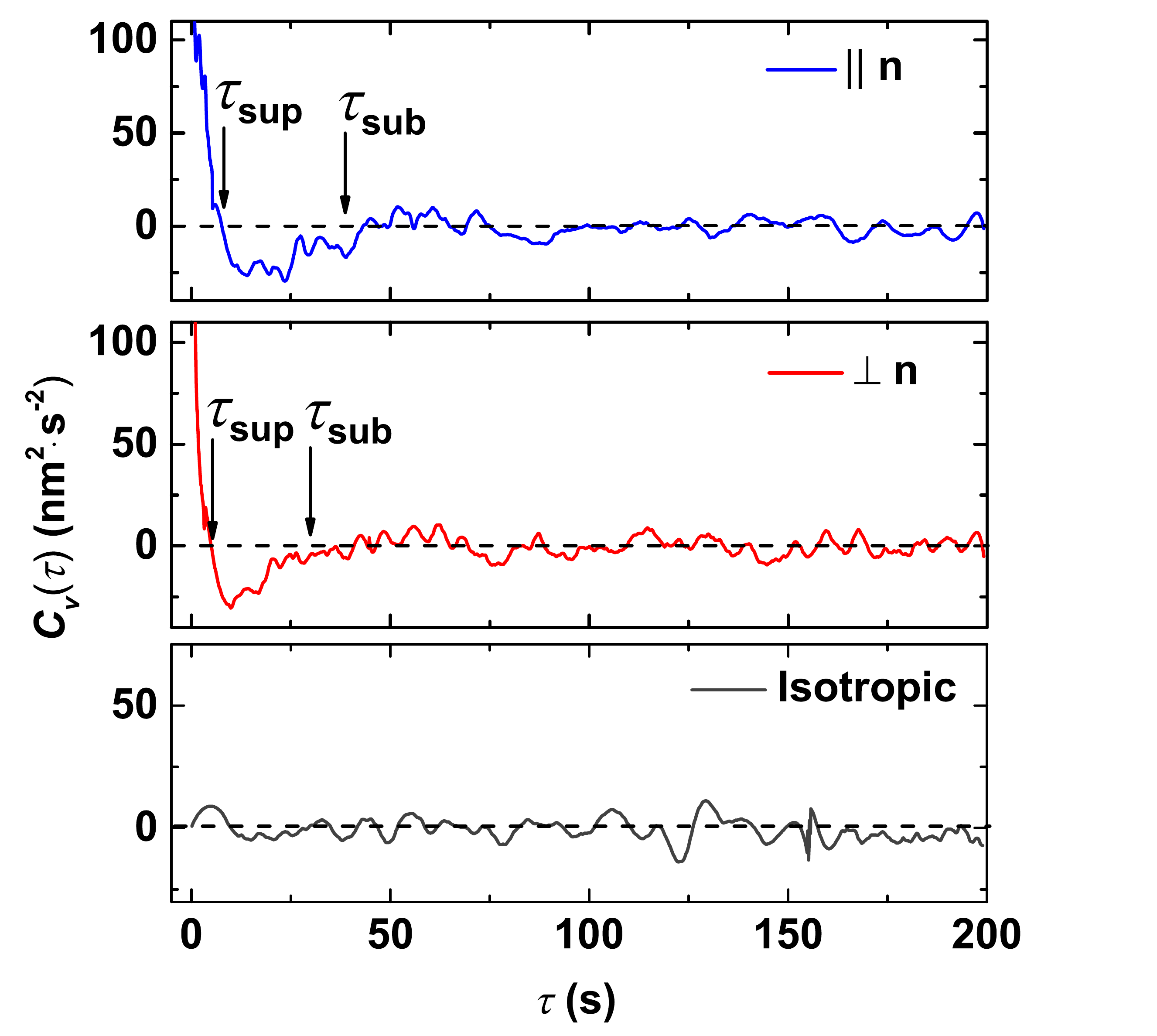}%
\\%
\parbox[t]{0.4\textwidth}{%
\centerline{(a)}%
}%
\hfill
\parbox[t]{0.4\textwidth}{%
\centerline{(b)}%
}
\caption{(Color online) (a) MSD versus time lag for 5~\textmu{}{m} silica particle with normal surface anchoring diffusing in the isotropic ($T=60$\textcelsius) and nematic ($T=50$\textcelsius) phases of IS--8200, in the directions parallel ($x$) and perpendicular ($y$) to the overall director. Cell thickness 50~\textmu{}{m}. (b) Velocity autocorrelation function calculated from experimental data for the normally anchored sphere moving in the nematic, parallel and perpendicular to the director and in isotropic phase. Arrows show $\tau _{\textrm{sup}} $ and $\tau _{\textrm{sub}} $, where $C_{v} \left(\tau \right)$ is close to~0.}
	\label{fig6}
\end{figure}

The measured MSD vs. time lag $\tau $  dependencies for $d=5$~\textmu{}m silica spheres in IS--8200 are presented in figure~\ref{fig6}~(a), for perpendicular anchoring. In the isotropic phase (elevated temperature $T=60$\textcelsius) the diffusion is normal in the entire range of time lags, with the diffusion coefficient $D=9.2\cdot 10^{-16}$~{m}$^{2}$/s. %
In the nematic, at $T=50$\textcelsius, the diffusion becomes anisotropic, with MSD being different when measured parallel and perpendicular to $\textbf{n}_{0} $. At relatively long time scales, $\tau >\left(20-40\right)$~{s}, both MSD components grow linearly with $\tau $, with the diffusion coefficients $D_{\parallel} =1.9\cdot 10^{-16}$~{m}$^2$/{s} and $D_{\bot } =1.4\cdot 10^{-16}$~{m}$^2$/{s} for the normally anchored spheres. At the times shorter than about (20--40)~s, the MSD time dependence becomes markedly nonlinear, figure~\ref{fig6}~(a). An apparent MSD of particles glued to the bottom of the cell (see figure~\ref{fig5}) is much smaller than the MSD in the nonlinear range and is practically time-independent; it merely adds a constant background to the MSD of the moving particles. Thus, the non-linear behavior at short times is not a spurious effect due to the finite accuracy of measurements \cite{Martin02} or birefringence.%

To obtain a better insight into the different diffusion regimes and the characteristic times limiting their borders, we calculated the velocity autocorrelation function $C_{v\parallel} (\tau )=\left\langle v_{x} (\tau )v_{x} (0)\right\rangle $ \cite{Cickochi95,Scher73}, where $v_{x} $ is the translational velocity of the particle along the $x$-axis, and a similar quantity  $C_{v\bot } (\tau )=\left\langle v_{y} (\tau )v_{y} (0)\right\rangle $ for $y$-direction. In fact, one calculates the autocorrelation function of the mean velocity over finite time interval between the position of the particle (time lag) which is much shorter than the correlation time. Under this condition, the mean velocity autocorrelation serves as a good estimate of the velocity autocorrelation function \cite{Brodin14}.%

For the diffusion in the isotropic fluid and for the diffusion at large time scales in the nematic, $C_{v\parallel } \left(\tau \right)$ and $C_{v\bot } \left(\tau \right)$ are  close to zero, figure~\ref{fig6}~(b), as it should be for the normal diffusion with MSD growing linearly with $\tau $. However, both $C_{v\parallel } \left(\tau \right)$ and $C_{v\bot } \left(\tau \right)$ become negative in the nematic, when the time lag is within a certain interval $\tau _{\sup } <\tau <\tau _{\textrm{sub}} $, indicating $\alpha <1$, i.e., subdiffusion. At shorter time scales $\tau <\tau _{\sup } $, $C_{v\parallel } \left(\tau \right)$ and $C_{v\bot } \left(\tau \right)$ become positive, indicating $\alpha >1$ and superdiffusion. Determination of $\tau _{\sup } $ is straightforward, as a point where the velocity autocorrelation functions change their sign. The value $\tau _{\textrm{sub}} $ is determined approximately when the deviation from zero exceeds $10^{-17} \; \textrm{m}^{2}/ \textrm{s}^{2} $, a typical scatter of VACF data in figure~\ref{fig6}~(b) as it was proposed in \cite{Turiv13}.%

\section{Discussion} \label{discussion}%

The anomalous Brownian motion is only observed in the nematic phase. As soon as the nematic is brought into isotropic state, the sub- and superdiffusive behavior changes to a normal linear MSD time dependence down to the shortest experimentally accessible times. Thus, anomalous dynamics must be related to the dynamics of additional degrees of freedom that exist in the nematic, but not in the isotropic phase, namely to the nematic director dynamics. The experiments demonstrate that the orientationally ordered environment influences the Brownian motion of a particle most profoundly, causing, in addition to anisotropy, anomalous super- and subdiffusion. The corresponding times, $\tau _{\textrm{sub}} $ and $\tau _{\textrm{sup}} $, vary with the type of anchoring at the particle's surface, size, and displacement direction \cite{Turiv13}. Above these time scales, the diffusion becomes normal (but still anisotropic). The anomalous character of the diffusion in the range $\tau <\tau _{\textrm{sub}} $ is evident not only in the nonlinear dependency of MSD on $\tau $, figure~\ref{fig6}~(a),  but also in the behavior of velocity correlation functions, figure~\ref{fig6}~(b).%

The current models of Brownian motion in a nematic \cite{Ruhnwald96,Mondiot12,Stark02,Stark03} consider the director field around the particle as being stationary. The predicted diffusion is always normal albeit anisotropic. Our results agree well with these models if $\tau $ is large, $\tau >\tau _{\textrm{sub}} $. At $\tau <\tau _{\textrm{sub}} $, however, the diffusion becomes anomalous; we attribute the effect to the director fluctuations.%

The relaxation times of director fluctuations relevant to the Brownian motion are expected to be limited by the interval $\tau _{d} <\tau <\tau _{h} $, where $\tau _{d} =\eta d^{2} /\pi ^{2} K$ and $\tau _{h} =\eta h^{2} /\pi ^{2} K$ are the characteristic times associated with the particle diameter and the cell thickness, respectively. At times $\tau >\tau _{h} $, the fluctuations are suppressed by the surface anchoring at the cell plates.  For $\tau \ll\tau _{d} $, the influence of perturbations with the wavelength much shorter than $d$ averages to zero over  the distance $d$. For $d=5$~\textmu{}{m} in an IS--8200 cell of thickness 50~\textmu{}{m}, the limits are estimated as $\tau _{d} \approx 0.3$~{s}  and $\tau _{h} \approx 30$~{s}. The time range $\tau _{d} <\tau <\tau _{h} $ thus embraces the experimentally determined $\tau _{\textrm{sub}} =\left(20-42\right)$~{s}  and $\tau _{\textrm{sup}} =\left(4-10\right)$~{s}, see figure~\ref{fig6}.%

The nematic is a viscoelastic medium, in which the director field $\mathbf{n} (\mathbf{r}, t)$ is coupled to the the velocity field $\mathbf{v} (\mathbf{r}, t)$. Both  $\mathbf{n} (\mathbf{r}, t)$ and $\mathbf{v} (\mathbf{r}, t)$ are perturbed by the particle and by the director fluctuations. Translational motion is coupled to the orientational dynamics of $\mathbf{n} (\mathbf{r}, t)$. In its turn, director reorientations induce torques and forces that cause the material to flow (the so-called backflow effect \cite{Kleman03, Pieranski73, Sunarso08}) and thus modify $\mathbf{v} (\mathbf{r}, t)$. The director fluctuations establish an intrinsic memory at the scales $\tau _{d} <\tau <\tau _{h} $, which is typically much longer than the hydrodynamic memory time of the isotropic fluid \cite{deGennes93}. The intrinsic memory is known to cause both superdiffusion and subdiffussion, sometimes just by varying the parameters of the very same system \cite{Kumar10}. Below we present qualitative effects that help to understand the connection of director fluctuations to the intrinsic viscoelastic memory and anomalous diffusion in the interval $\tau _{d} <\tau <\tau _{h} $.%

The equation of motion for the director fluctuations $\delta \mathbf{n}$ in a bulk nematic liquid crystal reduces to the torque equation for viscous and elastic torques \cite{deGennes93},
\begin{equation} \label{eq5}
	\gamma _{1} \frac{\partial }{\partial t} \delta \mathbf{n}=K\nabla ^{2} \delta \mathbf{n},
\end{equation}
where $\gamma _{1} $ is the rotational viscosity and $K$ is the effective elastic constant. Performing Fourier-Laplace transform of equation \eqref{eq5} yields the dispersion relation with purely imaginary frequency $\omega =-\ri Kq^{2} /\gamma _{1} $, from which it follows that the fluctuation modes are overdamped and thus purely relaxational with the relaxation time $\tau _{\mathbf{q}} =\gamma _{1} /(Kq^{2} )$. Their power spectrum is then
\begin{equation} \label{eq6}
	I_{\mathbf{q}} \left(\omega \right)=\frac{k_{\textrm{B}} T}{\pi Kq^{2} } \frac{\tau _{\mathbf{q}} }{1+\left(\omega \tau _{\mathbf{q}} \right)^{2} }
\end{equation}
with the corresponding correlation function
\begin{equation} \label{eq7}
	C_{\mathbf{q}} \left(\tau \right)=\left\langle \delta \mathbf{n}_{-\mathbf{q}} \left(0\right)\delta \mathbf{n}_{\mathbf{q}} \left(\tau \right)\right\rangle =\frac{k_{\textrm{B}} T}{Kq^{2} } \re^{-\tau /\tau _{\mathbf{q}} } .
\end{equation}
Fluctuation amplitude is thus $\left\langle \left|\delta \mathbf{n}_{\mathbf{q}} \right|^{2} \right\rangle \equiv C_{\mathbf{q}} \left(0\right)=k_{\textrm{B}} T/\left(Kq^{2} \right)$.

If a nematic is confined in a flat cell, then, due to restrictions imposed by the boundary conditions and particle's size, only a discrete set of fluctuation wave vectors is allowed. Consequently, the relaxation times of different fluctuation modes depend on the size of the inclusion, cell thickness and anchoring conditions at the surfaces \cite{Lee84,Stallinga96}.%

A coupling between the LC director and particle dynamics may result in a variety of scenarios of a particle movement. A dipolar inclusion behaves as an elastic dipole with the dipole moment $P=aR^{2}$ ($a=2.04$, reference \cite{Lubensky98}) and, therefore, interacts with inhomogeneities of the director field that arise due to thermal fluctuations. The dipole energy is then $U=-4\pi KP \nabla \mathbf{n}$, so that the particle experiences a force $\mathbf{F}=-\nabla U=4\pi KP \nabla \left(\nabla \mathbf{n}\right)$. Neglecting the inertial effects, the particle then moves with a velocity $\mathbf{v}$ that is proportional to the force,
\begin{equation} \label{eq8}
	\textbf{v} =\frac{\textbf{F} }{6\pi \eta R} =\frac{2KP}{3\eta R} \nabla (\nabla \cdot \delta \mathbf{n}).
\end{equation}
Following the Fourier-transformation into reciprocal space, the director fluctuation component $\delta \mathbf{n}_{\mathbf{q}}$ with wave vector $\mathbf{q}$ results in a particle velocity component
\begin{equation} \label{eq9}
	\textbf{v} _{\mathbf{q}} =-\frac{2KP}{3\eta R} \mathbf{q}(\mathbf{q}\cdot \delta \mathbf{n}_{\mathbf{q}} ).
\end{equation}

Particle velocity autocorrelation function is then
\begin{equation} \label{eq10}
	C_{\textbf{v},\textbf{q}} (\tau )=\left\langle \textbf{v} _{-\textbf{q}} (0)\, \cdot \, \textbf{v} _{\textbf{q}} (\tau )\right\rangle =\left(\frac{2KP}{3\eta R} \right)^{2} q^{2} \left\langle \textbf{q}\cdot \delta \textbf{n}_{-\textbf{q}} (0)\, \, \, \textbf{q}\cdot \delta \textbf{n}_{\textbf{q}} (\tau )\right\rangle.
\end{equation}%

For the fluctuation mode that involves splay deformation, $\delta \textbf{n}$ is in the ($\textbf{n},\textbf{q}$) plane and perpendicular to \textbf{n} \cite{deGennes93}, so that $\textbf{q} \cdot \delta \textbf{n} =q\delta n\sin \theta $, where $\theta $ is the angle between \textbf{q} and \textbf{n}. Particle velocity autocorrelation function is then proportional to the director autocorrelation function of equation~\eqref{eq7},
\begin{equation} \label{eq11}
	C_{\textbf{v},\textbf{q}} (\tau )=\left(\frac{2KP}{3\eta R} \right)^{2} q^{4} \sin ^{2} \theta \left\langle \delta \textbf{n} _{-\textbf{q} } (0)\, \delta \textbf{n} _{\textbf{q} } (\tau )\right\rangle =Aq^{2} \sin ^{2} \theta \, \re^{-\tau /\tau _{\textbf{q} } } ,
\end{equation}%
where we substitutesd the director correlation function from equation \eqref{eq7} and absorbed all constant prefactors into $A=(2P/3\eta R)^{2} k_\textrm{B} TK$. The obtained equation describes the contribution of thermal director fluctuations with wavevector $\textbf{q}$ to the particle velocity autocorrelation function. To obtain a full velocity correlation function, the equation should be integrated over $\textbf{q}$. (Note that fluctuations with different $\textbf{q}$ are uncorrelated.) Integration should only involve fluctuations occurring on length scales large compared to the particle size $d=2R$, which corresponds to the wavenumbers smaller than $q_{d} =C/d$, where $C$ is a constant of the order one. (The high-${q}$ fluctuations with the correlation length smaller than the size of the particle will exert uncorrelated forces on different parts of the particle, averaging out to zero.) Thus,

\begin{equation} \label{eq12}
	C_{\textbf{v}} (\tau )=A\int _{q<q_{d} }q^{2} \sin ^{2} \theta \, \re^{-\tau /\tau _{\textbf{q}} } \mathrm{d}\mathbf{q},
\end{equation}%
where $\mathrm{d}\mathbf{q}$ is the volume element in $\textbf{q}$-space.

The MSD is expressed through the velocity autocorrelation function as follows \cite{Scher73,Berne66}:
\begin{equation} \label{eq13}
	\left \langle \Delta x^{2}(\tau ) \right \rangle = 2\int_{0}^{\tau } \rd t'\int_{0}^{t'} C_{v_{x}}(t) \mathrm{d}t .
\end{equation}%
Conversely, $C_{v_{x}}(\tau) = 1/2 \, \frac{\mathrm{d}^{2}}{\mathrm{d}\tau ^{2}}\, \left \langle \Delta x^{2}(\tau ) \right \rangle$. The coupling mechanism discussed above evidently leads to \linebreak $C_{v}(\tau )>0$, and, therefore, to superdiffusion. Thus, direct dipole coupling to the director fluctuation dynamics may be responsible for the superdiffusion that we observe.%

If the director field $\textbf{n}(\textbf{r})$ around the particle is bent by an external action or thermal agitation, it will exert a torque on the particle, proportional to the director rate of change $\dot{\textbf{n}}(t)$. In response, the particle will rotate with angular velocity $\omega \propto \dot{\textbf{n}}(t)$. Neglecting the inertial effects, a particle rotating with an angular velocity $\omega $ will also move with a velocity $v$ proportional to $\left| \omega \right| $, and thus to $\left| \dot{\textbf{n}}(t) \right|$, see, e.g., references~\cite{Stark03,Jakli08}. Thus, the particle velocity is coupled to the director fluctuations of the surrounding nematic, and thus the particle velocity autocorrelation function is proportional to the ($q$-dependent) director angular velocity correlation function,
\begin{equation} \label{eq14}
	C_{\textbf{v},\textbf{q}} (\tau )=c\left\langle \delta \dot{\textbf{n}}_{-\textbf{q}} (0)\, \, \delta \dot{\textbf{n}}_{\textbf{q}} (\tau )\right\rangle,
\end{equation}%
where $c$ is a coupling constat. One needs to integrate over $q$, noting as before that only fluctuations with $q < q_{d}$ are relevant. The particle velocity correlation function then becomes
\begin{equation} \label{eq15}
	C_{\textbf{v}} (\tau )=c\int _{q<q_{d} }\left\langle \delta \dot{\textbf{n}}_{-\textbf{q}} (0)\, \, \delta \dot{\textbf{n}}_{\textbf{q}} (\tau )\right\rangle \mathrm{d}\mathbf{q}.
\end{equation}%
Recalling that, for any mechanical property $A$ that is a function on the phase space of a classical many-particle system, there holds $\langle \dot{A} (0) \dot{A} (t)\rangle = -\frac{\mathrm{d}^{2}}{\mathrm{d}t^{2}}\left\langle A(0) A(t)\right\rangle$ \cite{Berne66}, the director angular velocity correlation function $C_{\dot {\textbf{n}},\textbf{q}} (\tau )=\left\langle \dot{\textbf{n}}_{-\textbf{q}} (0)\, \, \dot{\textbf{n}}_{\textbf{q}} (\tau )\right\rangle = -\frac{\mathrm{d}^{2}}{\mathrm{d}\tau ^{2}}C_{\textbf{n},\textbf{q}} (\tau )$, where $C_{\textbf{n},\textbf{q}} (\tau )$ is given by equation \eqref{eq7}, so that
\begin{equation} \label{eq16}
	C_{\dot {\textbf{n}},\textbf{q}} (\tau )=\left\langle \dot{\textbf{n}}_{-\textbf{q}} (0)\, \, \dot{\textbf{n}}_{\textbf{q}} (\tau )\right\rangle = -\frac{k_\textrm{B}T}{Kq^{2}\tau_{\textbf{q}}^2}\re^{-\tau /\tau _{\textbf{q}}}.
\end{equation}%
Clearly, the director angular velocity autocorrelation function  is negative. This is easy to understand in view of the fact that director fluctuations are only small angular excursions from the mean, so that if the director rotates in certain direction at a given instant of time, at a later time it should be rotating back, which means a negative angular velocity autocorrelation. At short times, however, the relation equation \eqref{eq16} is not valid, as the initial value of an autocorrelation function must be positive. At short times, the inertial effects cannot be neglected, and correct asymptotic behavior of $C_{\textbf{n},\textbf{q}} (\tau )$ is such that its initial second derivative is negative, so that $C_{\dot {\textbf{n}},\textbf{q}} (\tau )$ is positive at short times, as it should be.

From equations \eqref{eq15} and \eqref{eq16} it follows that, if the particle motion couples to the director rotations, then the particle velocity autocorrelation function becomes
\begin{equation} \label{eq17}
	C_{\textbf{v}} (\tau )=-B\int _{q<q_{d} }q^{2} \re^{-\tau /\tau _{\textrm{q}}} \mathrm{d}\mathbf{q},
\end{equation}%
where $B$ is a constant. Thus, $C_{\textbf{v}} (\tau )$ due to this coupling mechanism is negative, and corresponds to sub\-diffusion.%

The coupling mechanism between the director bending modes and particle diffusion, discussed above, is indirect in the sense that director reorientations are directly coupled to particle rotations, and the latter may then couple to translations. Another mechanism that would couple the director bending modes with particle translation is through the backflow effects \cite{Oswald05}, whereby director fluctuations induce flows that affect the embedded particles. In the simplest realization in two dimensions, the force exerted on the fluid and that generates a backflow depends on the director angular velocity $\dot{\textbf{n}}$, its gradient $\nabla \dot {\textbf{n}}$, and the director field gradient \cite{Svensek01}. This force generates a viscose flow with a velocity $\textbf{v}$ proportional to the force, and thus depends on $\dot{\textbf{n}}$ and $\nabla \dot {\textbf{n}}$. Assuming that an embedded particle just follows the flow, its velocity autocorrelation function will thus depend on the director angular velocity autocorrelation function, which leads to subdiffusion, as discussed in the preceding paragraph. This is most readily seen for one of the contributions to the backflow, which is proportional to the projection of the director angular velocity gradient onto the director $\textbf{n}$, i.e., $\textbf{n}\nabla \dot {\textbf{n}}$ \cite{Svensek01}. The corresponding contribution to the particle velocity is then $\textbf{v} = k\textbf {n} \nabla \dot {\textbf{n}}$, where $k$ is a proportionality factor. Transforming to the reciprocal space,
\begin{equation} \label{eq18}
	v_{\textbf{q}}=k\left( \textbf{n} \cdot \ri \textbf{q} \right) \dot {n}_{\textbf{q}}=\ri kq\cos\theta \,\dot {n}_{\textbf{q}}\,,
\end{equation}%
where $\textbf{q}$ is the wavevector and $\theta$ is the angle between $\textbf{q}$ and $\textbf{n}$. Velocity autocorrelation function is then
\begin{equation} \label{eq19}
	C_{v,{\textbf{q}}}(\tau )=\left\langle v_{-\textbf{q}}(0) v_{\textbf{q}} (\tau ) \right\rangle=k^{2}q^{2} \left\langle \dot{n}_{-\textbf{q}}(0) \dot{n}_{\textbf{q}}(\tau ) \right \rangle.
\end{equation}%
Substituting $\left\langle \dot{n}_{-\textbf{q}}(0) \dot{n}_{\textbf{q}}(\tau ) \right \rangle$ from equation \eqref{eq16} and integrating over $\textbf{q}$ to obtain $C_{v}(\tau )$, one ends up with the same integral as in equation \eqref{eq17} (of course with a different pre-factor). Thus, the coupling to the director fluctuations through backflow effects leads to a negative particle velocity autocorrelation function of the form of equation \eqref{eq17}, and thus to subdiffusion.

The described mechanisms and their interplay may lead to intricate scenarios of coupled particle dynamics, depending on the time scales and relative strength of the two coupling effects. The time scales may in fact be well separated. Indeed, there are two independent director fluctuation modes that involve different types of director distortions \cite{deGennes93}. If the elastic constants pertaining to these deformations are much different, then these two modes will relax on different time scales. In particular, only one of the fluctuation modes involves a splay deformation, which is the only distortion that directly couples to elastic dipoles through the first mechanism discussed above.%
\begin{figure}[!h]
	\centering
		\includegraphics[scale=0.3, trim=0 0 0 50, clip]{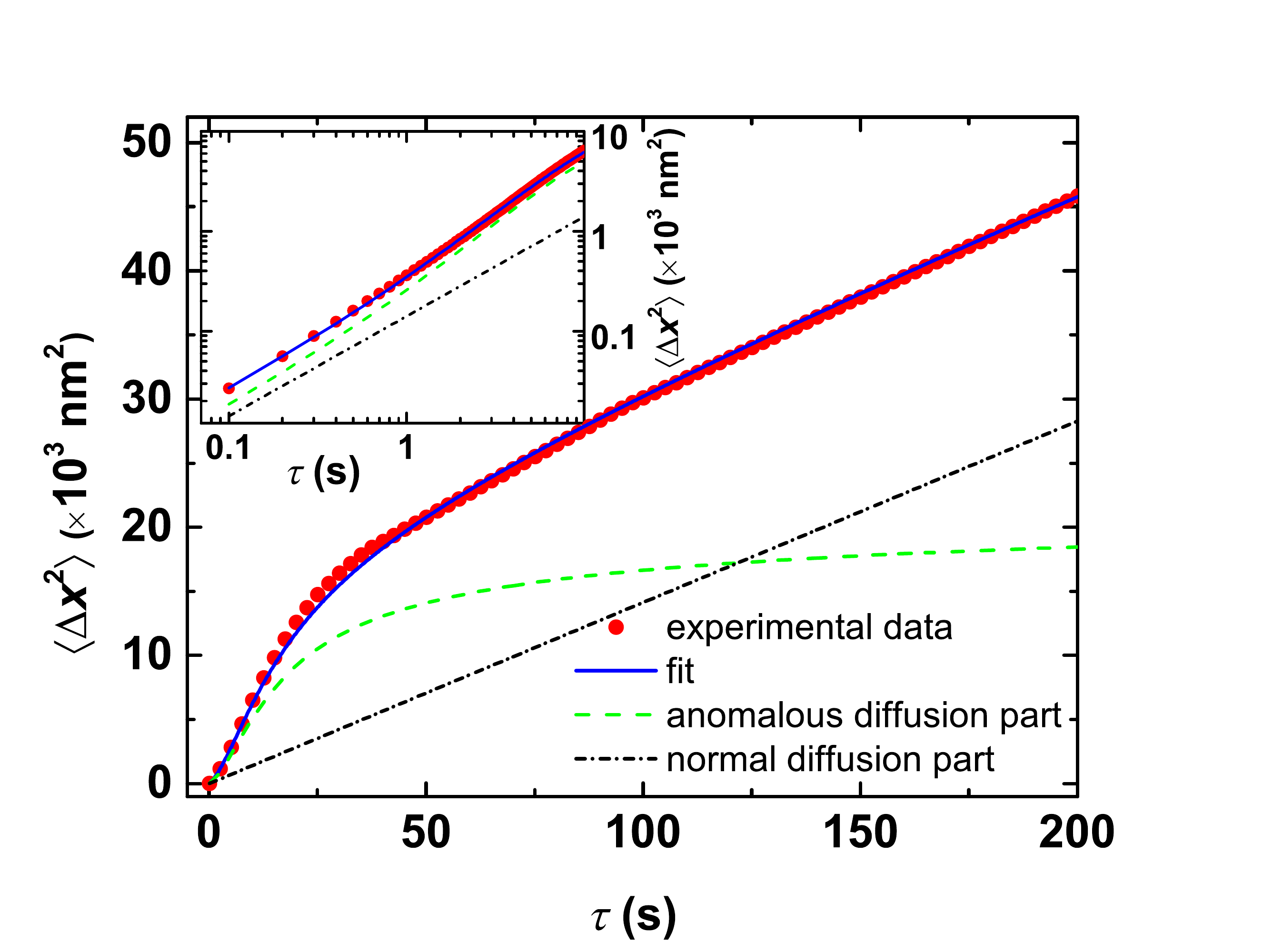}
			\caption{(Color online) MSD parallel to the nematic director for 5~\textmu{}m particle in IS--8200 along with theoretical curves for anomalous (green dash line) and normal contributions (black dash line) as well as fit (blue solid line) of the experimental data obtained as a combination of the previous two. Inset: data in double logarithmic scale in order to show fittings for short time scales.}
	\label{fig7}
\end{figure}%

The diffusion dynamics shown in figure~\ref{fig6} is a complex and multidimensional process that involves different factors and mechanisms. This complexity does not allow one to take into account all aspects of the diffusion process; however, our analysis gives a physical picture where the elastic interaction between the embedded particles and director fluctuations may describe the complex behavior of particle's dynamics in a NLC.

Figure~\ref{fig7} shows the experimental data of IS--8200 for MSD parallel to the nematic director, together with theoretical curves. The data are the same as in figure~\ref{fig6}~(a), with the background due to position determination errors (figure~\ref{fig5}) subtracted. In order to generate the curves in the figure, we used the viscosity $\eta=2.5$~{Pa}$\cdot${s} (manufacturer supplied data) and the elastic constants $K_{11}=K_{33}=30$~{pN} and $K_{22}=10$~{pN} (our estimates), from which the average elastic constants are $K_{1}=\frac{1}{3}K_{33}+ \frac{2}{3}K_{11}=30$~{pN} for splay (super\-diffusion) and $K_{2}=\frac{1}{3}K_{33}+ \frac{2}{3}K_{22}\approx 17$~{pN} for bend (subdiffusion) \cite{deGennes93}. Contributions of the super- and sub\-diffusion to MSD were computed by integration, equation \eqref{eq13}, of the velocity correlation functions of equation \eqref{eq12} and \eqref{eq17}, respectively. We assumed that $q_{d}=C/d$, where $C$ is a constant of the order of one. To obtain the fit in the figure, we varied $C$ and the magnitudes of the super- and subdiffusive contributions. Fit results are seen to agree well with the experimental data, both in the regions of super- (inset in figure~\ref{fig7}) and subdiffusion (figure~\ref{fig7}).

\section{Conclusions}\label{conclusions}

We investigate the thermal motion of colloidal particles in a nematic liquid crystal for the time scales shorter than the expected time of director fluctuations. At long times, compared to the characteristic time of the nematic director fluctuations, we observe a typical anisotropic Brownian motion with the MSD linear in time and inversely proportional to the effective viscosity of the nematic medium. At shorter times, however, the dynamics is markedly nonlinear and exhibits subdiffusive (slower than $\propto \tau $) evolution of MSD. We present a simple illustration of how the director fluctuations influence the Brownian motion through the long-range interactions. Although this study dealt with standard liquid crystals, the observed anomalous diffusion is expected to arise in any dispersive environment with reduced symmetry and orientational order. This is a remarkable finding, since local orientational order should profoundly influence the dynamics of many complex biological systems, such as cell membranes, the cycloskeleton, assembles of anisotropic particles, etc. A detailed theoretical description of the effect is highly desirable.

\section*{Acknowledgements}

We thank Lavrentovich~O.D., Lev~B.I., Lazo~I., Soroka~P., Pergamenshchyk~V., Chernyshuk~S. and Boiko~O. for fruitful discussions. The article was supported by NASU project No.~1.4.B/162.

\ukrainianpart \label{ukr}
\title{Аномальний броунівський рух колоїдної частинки в нематичному середовищі: вплив флуктуацій директора}
\author{Т. Турів\refaddr{label1, label2}, О. Бродин\refaddr{label1, label3}, В.Г. Назаренко\refaddr{label1}} %
\addresses{
\addr{label1} Інститут фізики НАН України,  просп. Науки, 46,  03028 Київ, Україна
\addr{label2} Інститут рідких кристалів, Кентський державний університет,  вул. Університетська Еспланада, 1425,  44242 Кент, Огайо, США
\addr{label3} Національний технічний університет України ``КПІ'',  просп. Перемоги, 37, 03056 Київ, Україна
} %
\makeukrtitle
\begin{abstract}
\tolerance=3000%
Як було нещодавно опубліковано в [Turiv~T. et al., Science, 2013, \textbf{342}, 1351], флуктуації в орієнтації рідкокристалічного (РК) директора можуть переносити імпульс від РК до колоїда, тоді дифузія самого колоїда стає аномальною на коротких проміжках часу. Використовуючи методи відеомікроскопії та одночастинкового відстеження, ми досліджували випадковий тепловий рух колоїдної частинки в нематичному РК на часах, коротших за очікуваний час флуктуацій директора. На довгих часах, в порівняні з характеристичним часом релаксації нематичного директора, ми спостерігали типовий анізотропний броунівський рух з середнім квадратом зміщення частинки пропорційним до часу $\tau $ та обернено пропорційним до ефективної в'язкості нематичного середовища. Однак на коротших часах, динаміка помітно відрізняється від лінійної, середній квадрат зміщення в цьому випадку зростає повільніше (субдифузія) або швидше (супердифузія) від $\tau $. Результати пояснюються в контектсі зв'язку динаміки колоїдної частинки з динамікою флуктуацій директора.
\keywords нематичний рідкий кристал, броунівський рух, флуктуації директора%
\end{abstract}
\end{document}